\documentclass[twocolumn,showpacs,preprintnumbers,amsmath,amssymb,groupedaddress,pre]{revtex4}

\usepackage{graphicx}
\usepackage{dcolumn}
\usepackage{bm}
\usepackage{epsfig}

\def\e{\begin{equation}}
\def\f{\end{equation}}
\def\=#1{\overline{\overline #1}}

\def\-#1{{\bf #1}}
\def\.{\cdot}

\def\r#1{(\ref{eq:#1})}
\def\vec#1{{\bf #1}}

\begin{document}

\title{Resolution of sub-wavelength transmission devices formed by a wire medium}

\author{Pavel A. Belov}
\affiliation{Department of Electronic Engineering, Queen Mary
University of London, Mile End Road, London, E1 4NS, United Kingdom}
\affiliation{Photonics and Optoinformatics Department, St.
Petersburg State University of Information Technologies, Mechanics
and Optics, Sablinskaya 14, 197101, St. Petersburg, Russia}

\author{M\'ario G. Silveirinha}
\affiliation{Departamento de Eng. Electrot\'{e}cnica da Universidade
de Coimbra, Instituto de Telecomunica\c{c}\~{o}es, P\'{o}lo II, 3030
Coimbra, Portugal}

\begin{abstract}
The restrictions on the resolution of transmission devices formed by
wire media (arrays of conductive cylinders) recently proposed in
[Phys. Rev. B, 71, 193105 (2005)] and experimentally tested in
[Phys. Rev. B, 73, 033108 (2006)] are studied in this paper using
both analytical and numerical modeling. It is demonstrated that such
transmission devices have sub-wavelength resolution that can in
principle be made as fine as required by a specific application by
controlling the lattice constant of the wire medium. This confirms
that slabs of the wire medium are unique imaging devices at the
microwave frequency range, and are capable of transmitting
distributions of TM-polarized electric fields with nearly unlimited
sub-wavelength resolution to practically arbitrary distances.
\end{abstract}

\maketitle

\section{Introduction}
The resolution of common imaging systems is restricted by the
so-called diffraction limit, since these systems operate only with
propagating spatial harmonics emitted by the source. The
conventional lenses can not transport evanescent harmonics which
carry sub-wavelength information, since these waves exhibit
exponential decay in natural materials and even in free space.
Recently, a new type of devices capable of transmitting images with
sub-wavelength resolution was suggested in \cite{canal}. These
structures are formed by planar slabs of materials with specific
electromagnetic properties which make possible to transfer images
from one interface to another with sub-wavelength resolution.  Such
regime of operation is called canalization. This regime dramatically
differs from the sub-wavelength imaging principle formulated in
\cite{Pendrylens} which employs effects of negative refraction and
amplification of evanescent waves. The idea is that both propagating
and evanescent harmonics of a source can be transformed into the
propagating waves inside of a slab of an electromagnetic crystal.
Then, these propagating modes can transmit sub-wavelength images
from one interface of the slab to the other one.

\begin{figure}[h]
\centering \epsfig{file=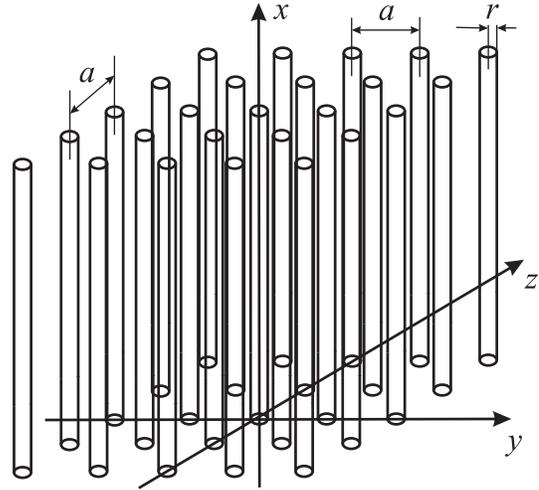, width=7cm} \caption{The
geometry of the wire medium: a square lattice of parallel ideally
conducting thin wires.} \label{geom}
\end{figure}

In order to implement the canalization regime it is required to use
materials in which the electromagnetic waves have a flat
isofrequency contour. Such media are capable of transmitting energy
only in one direction, always with the same phase velocity. The
typical example of such materials is the wire medium (see Fig.
\ref{geom}), which consists of an array of parallel ideally
conducting wires \cite{Rotmanps,Brown,pendryw,WMPRB}, operating at
frequencies smaller than its characteristic plasma frequency.
The wire medium supports transmission line modes which travel along
the wires at the speed of light. A slab of wire medium with
thickness equal to an integer number of half-wavelengths experiences
Fabry-Perot resonance for any angle of incidence, including complex
ones, and thus such a slab is capable of transporting images with
sub-wavelength resolution \cite{SWIWM}. The transmission devices
formed by wire medium were studied both numerically and
experimentally in \cite{SWIWM}. As result, the imaging with
$\lambda/15$ resolution has been successfully demonstrated. In
\cite{SWIWM} it was assumed that the structure operates at a very
low frequency as compared to its plasma frequency and no estimation
has been made on how the plasma-like properties of the wire medium
could affect the operation of such devices. In order to reveal the
restrictions on the resolution caused by plasma-like properties of
the wire medium, in the present paper we theoretically study the
transmission through a wire medium slab using the additional
boundary condition recently formulated in \cite{MarioABC}.  The wire
medium is modeled as a homogeneous material described by an
effective permittivity tensor. The theoretical results are validated
using the full wave periodic method of moments \cite{Wu}, which
models every detail of the structure and the actual granularity of
the material. It is proved that the effective material model is
sufficiently accurate to model the electromagnetic response of the
wire medium.

The paper is organized as follows. After a brief introduction, the
expressions of the reflection and transmission coefficients for a
planar slab of wire medium are presented. In the third section, the
transmission and reflection properties are studied for various
frequencies of operation using both theoretical and numerical
modeling. The fourth section is devoted to the study of resolution
using the Rayleigh criterion. The imaging accuracy and the
theoretical limit of resolution are studied in the fifth section
using a half-intensity criterion. Finally, the conclusion is
presented. Also, the paper is accompanied by an appendix where
guided modes in slabs of wire medium are investigated.

\section{Reflection and transmission coefficients for a wire medium slab}

The wire medium is a material characterized by strong spatial
dispersion even at low frequencies \cite{WMPRB}. It can be described
by a spatially dispersive permittivity tensor of the form: \e
\=\varepsilon=\varepsilon \-x\-x+\-y\-y+\-z\-z,\ \varepsilon
(\omega,q_x)=1-\frac{k_p^2}{k^2-q_x^2}, \label{eq:eff}\f where the
$x$-axis is oriented along wires, $q_x$ is the $x$-component of wave
vector $\vec q$, $k=\omega/c$ is the wave number of the host medium,
$c$ is the speed of light, and $k_p=\omega_p/c$ is the wave number
corresponding to the plasma frequency $\omega_p$. The plasma
frequency depends on the lattice period $a$ and on the radius of
wires $r$ \cite{JEWAwm}: \e k_p^2=\frac{2\pi/a^2}{\ln\frac{a}{2\pi
r}+0.5275}. \label{eq:k0}\f Note that in this paper we assume that
the metallic wires are arranged in a square lattice. The
permittivity tensor is normalized to the permittivity of the host
medium.

The solution of any boundary problem involving the wire medium
(using an effective medium theory) requires the knowledge of an
additional boundary condition at the interface. In fact, it has long
been known that the usual boundary conditions (continuity of the
tangential components of the electromagnetic field) are insufficient
in case of spatial dispersion \cite{Agranovich,Agarwal,Birman}. Such
an additional boundary condition has been formulated in
\cite{MarioABC} for the wire medium case. It was proved that the
normal component of the electric field must be continuous at the
interface between the wire medium and free space, under the
condition that the host medium of the wire medium is also free
space. In this paper, we will use this result to study the
resolution of the transmission devices formed by the wire medium.
Firstly, we characterize the scattering of plane waves by a wire
medium slab.

\begin{figure}[h]
\centering \epsfig{file=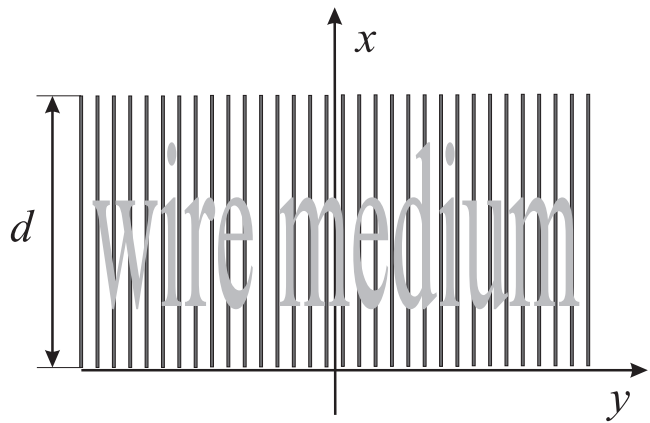, width=8.5cm} \caption{The wire
medium slab. The structure is unbounded and periodic along the $y$-
and $z$-directions.} \label{slab}
\end{figure}

Let us consider a slab of wire medium with thickness $d$ (see Fig.
\ref{slab}). We suppose that the wires stand in free-space and are
normal to the interface. The wires are assumed perfectly conducting,
i.e. losses are assumed negligible. The distribution of the
electromagnetic field at the input plane of the transmission device
can be decomposed in terms of spatial harmonics. The spatial
harmonics are either propagating plane waves or evanescent plane
waves, and their polarization can be classified as transverse
electric (TE) or transverse magnetic (TM) with respect to the
orientation of the wires. In \cite{SWIWM} it was proved that a slab
of the wire medium allows sub-wavelength imaging of the part of the
spatial spectrum with TM-polarization.

To determine the resolution of the imaging, we consider that a plane
wave with TM polarization impinges on the slab. Let $H_{\rm inc}$ be
the amplitude of the incident magnetic field along the
$z$-direction, and let $\vec k=(k_x,k_y,0)^T$ be the wave vector of
the incident wave. The $x$-component of the wave vector $k_x$ can be
expressed in terms of the wave number $k$ and of the $y$-component
of the wave vector $k_y$ as $k_x=-j\sqrt{k_y^2-k^2}$. The incident
plane wave excites two types of waves in the wire medium: the
extraordinary wave (TM mode) and the transmission line (TEM) mode
\cite{WMPRB}. The ordinary wave is not excited since it has TE
polarization. The tangential component of the wave vector $k_y$ is
preserved at the interface between free space and the wire medium.
Using this property we can evaluate the wave vector associated with
each of the excited modes in the slab. The wave vector associated
with transmission line modes is of the form: \e \vec q_\pm^{\rm
TEM}=(\pm k,k_y,0)^T. \f These waves travel along the wires with the
speed of light in free space, independently of the value of
transverse component of wave vector $k_y$. On the other hand, the
extraordinary modes obey to the dispersion equation derived in
\cite{WMPRB}: \e q_x^2+q_y^2+q_z^2=k^2-k_p^2, \f and thus the wave
vector of an extraordinary mode is of the form: \e \vec q_\pm^{\rm
TM}=(\pm q_x,k_y,0), \quad q_x=-j \sqrt{k_p^2+k_y^2-k^2}. \f

If the frequency of operation is smaller then the plasma frequency
($k<k_p$), then the extraordinary modes are evanescent: the
$x$-component of wave vector $q_x$ is a purely imaginary number for
any real $k_y$. In order to emphasize this fact it is convenient to
use the notation $q_x=-j\gamma_{\rm TM}$ with $\gamma_{\rm
TM}=\sqrt{k_p^2+k_y^2-k^2}$. Also, by analogy, we denote
$k_x=-j\gamma_x$ with $\gamma_x=\sqrt{k_y^2-k^2}$. The value of
$\gamma_x$ is purely imaginary if $k_y<k$ (propagating modes in free
space), but it becomes a real number if $k_y>k$ (evanescent modes in
free space, i.e. the sub-wavelength spatial spectrum).

The total magnetic field (directed along the $z$-axis) can be
written in the following form:
 \e \frac{H(x)}{H_{\rm inc}} =\left\{
\begin{array}{lcl}
e^{-jk_xx} + R e^{jk_xx},\quad x<0 \\[2mm]
\begin{array}{lcl}
A_-^{\rm TM}e^{-\gamma_{\rm TM}(x-d/2)}\\
+ A_+^{\rm TM}e^{+\gamma_{\rm TM}(x-d/2)}\\
+A_-^{\rm TEM}e^{-jk(x-d/2)}\\
+ A_+^{\rm TEM}e^{+jk(x-d/2)},\\
\end{array} \quad 0\le x\le d \\[9mm]
T e^{-jk_x(x-d)},\quad x>d, \vphantom{\frac{A}{A}}\\
\end{array}\right.
\label{eq:H} \f where $R$ and $T$ are the reflection and
transmission coefficients, $A_\pm^{\rm TM}$ and $A_\pm^{\rm TEM}$
are the amplitudes the extraordinary (TM) and transmission-line
(TEM) modes in the slab, respectively.

Following \cite{MarioABC}, all components of both magnetic and
electric fields are continuous at the interface between free space
and the wire medium. Thus, the magnetic field $H(x)$ and both its
first $dH(x)/dx$ and second $d^2H(x)/dx^2$ derivatives by $x$ are
continuous at interfaces $x=0$ and $x=d$. This can be written using
\r{H} in the form of the following system of equations:

\begin{widetext}

\e \left(
  \begin{array}{cccccc}
    -1 & e^{\gamma_{\rm TM}d/2} & e^{-\gamma_{\rm TM}d/2} & e^{jkd/2} & e^{-jkd/2} & 0 \\
    -jk_x & -\gamma_{\rm TM}e^{\gamma_{\rm TM}d/2} & \gamma_{\rm TM}e^{-\gamma_{\rm TM}d/2} & -jke^{jkd/2} & jke^{-jkd/2} & 0 \\
    k_x^2 & \gamma_{\rm TM}^2e^{\gamma_{\rm TM}d/2} & \gamma_{\rm TM}^2e^{-\gamma_{\rm TM}d/2} & -k^2e^{jkd/2} & -k^2e^{-jkd/2} & 0 \\
    0 & e^{-\gamma_{\rm TM}d/2} & e^{\gamma_{\rm TM}d/2} & e^{-jkd/2} & e^{jkd/2} & -1\\
    0 & -\gamma_{\rm TM}e^{-\gamma_{\rm TM}d/2} & \gamma_{\rm TM}e^{\gamma_{\rm TM}d/2} & -jke^{-jkd/2} & jke^{jkd/2} & jk_x\\
    0 & \gamma_{\rm TM}^2e^{-\gamma_{\rm TM}d/2} & \gamma_{\rm TM}^2e^{\gamma_{\rm TM}d/2} & -k^2e^{-jkd/2} & -k^2e^{jkd/2} & k_x^2\\
  \end{array}
\right) \left(
  \begin{array}{c}
    R\\
    A_-^{\rm TM}\\
    A_+^{\rm TM}\\
    A_-^{\rm TEM}\\
    A_+^{\rm TEM}\\
    T\\
  \end{array}
\right) = \left(
  \begin{array}{c}
    1\\
    -jk_x\\
    -k_x^2\\
    0\\
    0\\
    0\\
  \end{array}
\right) \label{eq:syst}\f

Solving this system the reflection and transmission coefficients can
be expressed as follows:

\e R=1-\frac{1}{1+\frac{\gamma_{\rm
TM}k_y^2}{\gamma_x(k_y^2+k_p^2)}\tanh(\frac{\gamma_{\rm TM}d}{2})-
\frac{kk_p^2}{\gamma_x(k_y^2+k_p^2)}\tan(\frac{kd}{2})} -
\frac{1}{1+\frac{\gamma_{\rm TM}k_y^2}{\gamma_x(k_y^2+k_p^2)}{\rm
ctanh}(\frac{\gamma_{\rm
TM}d}{2})+\frac{kk_p^2}{\gamma_x(k_y^2+k_p^2)}{\rm
ctan}(\frac{kd}{2})}, \label{eq:R} \f

\e T=\frac{1}{1+\frac{\gamma_{\rm
TM}k_y^2}{\gamma_x(k_y^2+k_p^2)}\tanh(\frac{\gamma_{\rm TM}d}{2})-
\frac{kk_p^2}{\gamma_x(k_y^2+k_p^2)}\tan(\frac{kd}{2})} -
\frac{1}{1+\frac{\gamma_{\rm TM}k_y^2}{\gamma_x(k_y^2+k_p^2)}{\rm
ctanh}(\frac{\gamma_{\rm
TM}d}{2})+\frac{kk_p^2}{\gamma_x(k_y^2+k_p^2)}{\rm
ctan}(\frac{kd}{2})}. \label{eq:T} \f
\end{widetext}

In the above "tanh" and "ctanh" represent the hyperbolic tangent and
cotangent, respectively. The canalization regime is observed for the
slab under consideration when $kd=n\pi$, where $n$ is an integer. In
this case \r{T} can be simplified: if $kd=(2n+1)\pi$ then \e T=-
\frac{1}{1+\frac{\gamma_{\rm TM}k_y^2}{\gamma_x(k_y^2+k_p^2)}{\rm
ctanh} (\gamma_{\rm TM}d/2)}, \quad R=1+T, \label{eq:t2n1}\f whereas
if $kd=2n\pi$ then \e T= \frac{1}{1+\frac{\gamma_{\rm
TM}k_y^2}{\gamma_x(k_y^2+k_p^2)}{\rm tanh} (\gamma_{\rm TM}d/2)},
\quad R=1-T. \label{eq:t2n}\f

The transmission coefficient calculated above can be regarded as a
transfer function: the image at the output plane is obtained by
superimposing the (TM-polarized) spatial harmonics of the source
field distribution at the input plane multiplied by $T$. Strictly
speaking, that is not rigourously true, even if the exact expression
for transmission coefficient is known. Indeed, when doing this we
are neglecting the higher order harmonics, which decay much faster
than the fundamental mode, but also contribute to the near field.
These harmonics are a manifestation of the actual granularity of
wire medium, and are not described by the effective medium model
\r{eff}. The contribution of the high order harmonics can be
neglected while $k_ya<\pi$, which corresponds to operation below the
ultimate limit of resolution $\Delta_l=a$ \cite{Subwavelength}. In
this case, slight variations of the near field near the edges of the
wires do not influence the resolution of the device. Thus, in this
paper we will consider that the distribution of the field at the
output plane is equal to the distribution of the field at the input
plane multiplied by the transmission coefficient. Within this
approximation, it is sufficient to characterize $T$ in order to
evaluate the resolution of the transmission device.

If the operating frequency is much smaller than the plasma
frequency, i.e. $k \ll k_p$, then since $kd/\pi > 1$ we can assume
that $\gamma_{\rm TM}d \gg 1$, and using the expressions \r{t2n1}
and \r{t2n} we obtain simple approximate formulae for the reflection
and transmission coefficients: \e T\approx \mp
\frac{1}{1+\frac{\gamma_{\rm TM}k_y^2}{\gamma_x(k_y^2+k_p^2)}},
\quad R\approx 1-\frac{1}{1+\frac{\gamma_{\rm
TM}k_y^2}{\gamma_x(k_y^2+k_p^2)}}, \label{eq:app}\f where the $\mp $
sign corresponds to the case in which the thickness of the slab is
equal to an even or odd number of half-wavelengths, respectively.

Some important properties of \r{t2n1} and \r{t2n} are readily
identified. It is clear that if $k_p\to \infty$ then $T\approx \mp
1$ for any $k_y$, and therefore, in such conditions, the imaging is
perfect for this polarization. On the other hand, one can see that
$T = 0$ when $\gamma_x = 0$. i.e. $k_y = k$. This property is very
important because it shows that the spatial harmonics with $k_y=k$
are filtered by the WM, which some how contradicts the
sub-wavelength imaging. Actually, as it will be demonstrated below,
the band of the spatial harmonics corresponding to this filtering is
so narrow that it practically does not affect the imaging properties
of the system.

It is noteworthy that if $k_y\to \infty$ then $T\to \mp 1/2$
\footnote{Nevertheless one should keep in mind that the theoretical
model is no longer valid when $k_y a > \pi$.}. This means that the
spatial harmonics in the deep subwavelength spectrum are transferred
from the source to the image plane, but in contrast to the spatial
harmonics with smaller $k_y$, their amplitude will be reduced two
times. Thus, the transfer function of the imaging device is not
perfect since it is not a constant function of the transverse wave
vector. Nevertheless,
 it will be shown ahead that this fact does not
significantly affect the sub-wavelength imaging performance of the
device.

Also, it is clear that the imaging quality does not depend on the
thickness of the structure. The transmission device can be made as
thick as it may be required by an application. The only restriction
is that the thickness has to be equal to an integer number of
half-wavelengths. If the number of half-wavelengths is odd then the
image at the back interface of the structure will appear out-of
phase with respect to the source, whereas if the number of
half-wavelengths is even then the image and the source will be in
phase. Moreover, the bandwidth of operation does not depend on the
thickness of the transmission device. It remains the same for all
slab thicknesses equal to the integer number of half-wavelengths.

\section{Parametric study of reflection and transmission properties}

In \cite{SWIWM} the wire medium with period $a=1$~cm formed by wires
with $r=1$~mm has been considered. Substitution of these parameters
into \r{k0} gives $k_pa=2.5$. A more accurate value, calculated
using the numerical method proposed in \cite{MarioMTT}, is
$k_pa=2.36$. The frequency of operation for the transmission device
in \cite{SWIWM} was $1$GHz, which corresponds to $ka=0.2$, and the
thickness $d=15$~cm of the structure was chosen equal to
half-wavelength, that is $kd=\pi$, or in terms of the wave number
corresponding to the plasma frequency $k_pd/\pi=11.3$. In the
present paper we consider the same parameters for the transmission
device.

\begin{figure}[htb]
\centering \epsfig{file=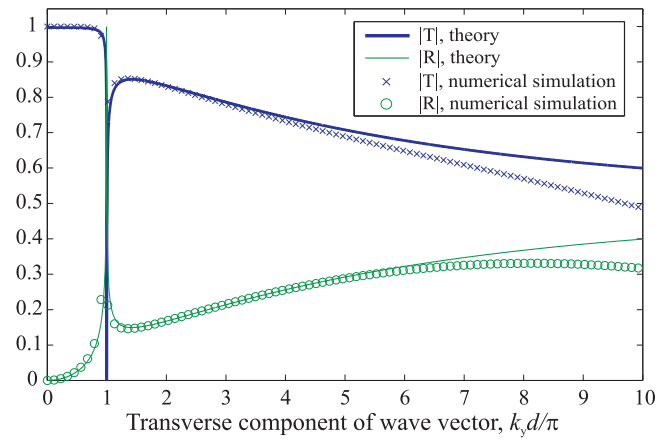, width=8.5cm} \caption{(Color
online) Reflection ($R$) and transmission ($T$) coefficients
(absolute values) as a function of the transverse component of wave
vector $k_yd/\pi$ for $kd/\pi=1$. Solid lines: analytical
expressions \r{R} and \r{T}. Crosses and circles: points calculated
using full wave numerical method.} \label{ka1}
\end{figure}

The dependence of the reflection and transmission coefficients
(absolute values) on the normalized transverse component of wave
vector $k_yd/\pi$ is plotted in Fig. \ref{ka1}. One can see that the
transmission coefficient has a very sharp dip at $k_y=k$, but for
other values of $k_y$ it has values close to unity. In Fig.
\ref{ka1} the "crosses" and "circles" represent the data calculated
numerically using the periodic method of moments. As seen, the
agreement between the analytical model and the full wave results is
very satisfactory. We do not present a plot for the phase of
transmission coefficient since it is practically equal to $\pi$ for
whole range of $k_y$, except for a very narrow band with upper bound
equal to $k$. The obtained dependence confirms that the slab indeed
operates in the canalization regime and is capable of transporting
sub-wavelength images with TM polarization.

\begin{figure}[htb]
\centering \epsfig{file=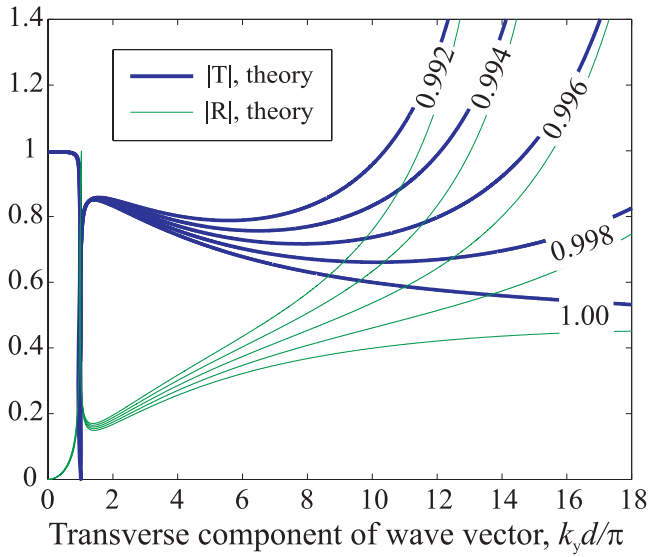, width=8.5cm} \caption{(Color
online) Reflection ($R$) and transmission ($T$) coefficients
(absolute values) as a function of the transverse component of wave
vector $k_yd/\pi$ for $kd/\pi=1-0.002n$, where n=0,1,2,3,4. The
numbers in the figure correspond to the values of $kd/\pi$.}
\label{m002468}
\end{figure}

The behavior of  the transmission and reflection coefficients is
sensitive to variations in the frequency of operation. In Fig.
\ref{m002468}, the transmission characteristic is depicted for
frequencies that are lower than original frequency of operation by
$0.2, 0.4, 0.6$ and $0.8$\%. It is seen that the amplitude of the
evanescent modes is amplified in a certain range of $k_y$.

\begin{figure}[htb]
\centering \epsfig{file=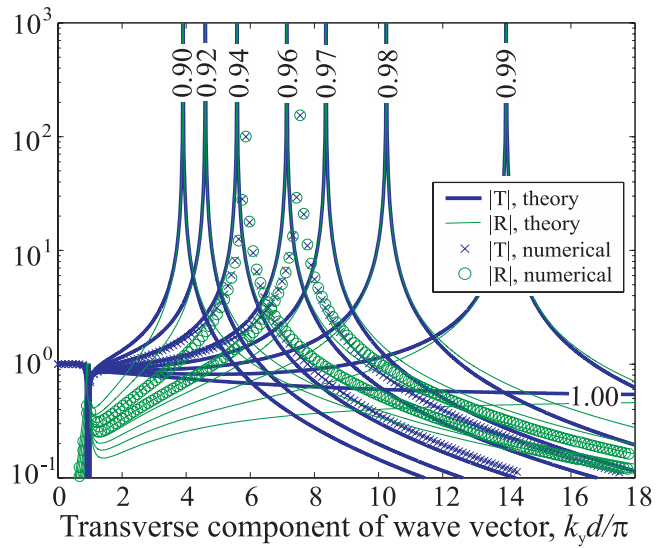, width=8.5cm} \caption{(Color
online) Reflection ($R$) and transmission ($T$) coefficients
(absolute values) as a function of the transverse component of wave
vector $k_yd/\pi$ for $kd/\pi=1-0.01n$, where n=0,1,2,3,4,6. The
numbers in the figure correspond to the values of $kd/\pi$. The
crosses and circles represent the results of the numerical modeling
for $kd/\pi=0.94$ and $kd/\pi=0.96$.} \label{m0123456}
\end{figure}

The further decrease of the frequency of operation reveals a very
interesting resonant phenomenon. This is illustrated in Fig.
\ref{m0123456},
where the transmission coefficient is plotted for
frequencies which lower than the design frequency by $1, 2, 3, 4$
and $6$\%. One can observe a very strong enhancement of certain
spatial harmonics. This property reveals the presence of guided
modes propagating along the $y$ direction of the slab, being the
mechanism of propagation closely related to that of an Yagi antenna
array \cite{Yagi}. The dispersion properties of the guided modes are
studied in the Appendix. The observed resonant enhancement
deteriorates the sub-wavelength imaging since some of the spatial
harmonics happen to be amplified by a factor of 10 or even more.

\begin{figure}[htb]
\centering \epsfig{file=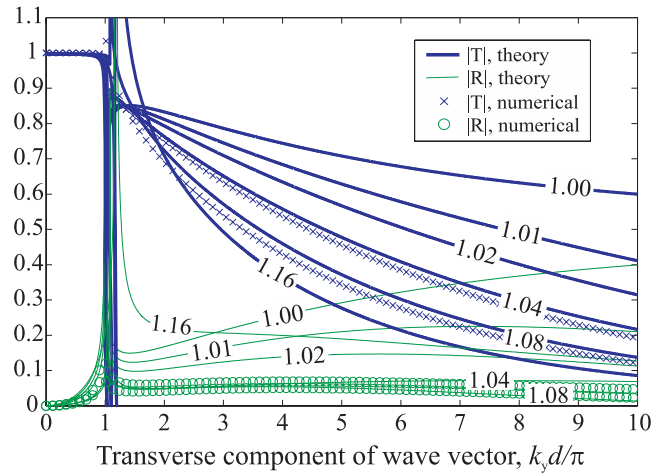, width=8.5cm} \caption{(Color
online) Reflection ($R$) and transmission ($T$) coefficients
(absolute values) as a function of the transverse component of wave
vector $k_yd/\pi$ for $kd/\pi=1+0.01n$, where n=0,1,2,4,8,16. The
numbers in the figure correspond to the values of $kd/\pi$. The
crosses and circles present the results of the numerical modeling
for $kd/\pi=1.04$ and $kd/\pi=1.08$.} \label{p01234}
\end{figure}
\begin{figure}[htb]
\centering \epsfig{file=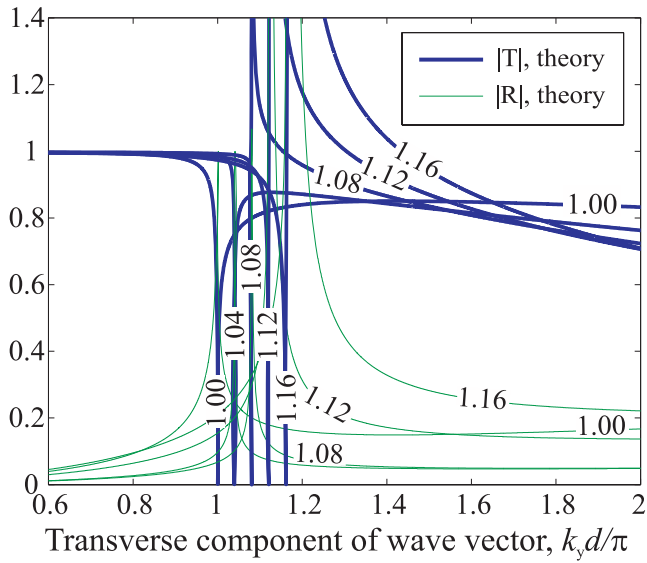, width=8.5cm} \caption{(Color
online) Reflection ($R$) and transmission ($T$) coefficients
(absolute values) as a function of the transverse component of wave
vector $k_yd/\pi$ for $kd/\pi=1+0.04n$, where n=0,1,2,3,4. The
numbers in the figure correspond to the values of $kd/\pi$}
\label{p0481216}
\end{figure}

For frequencies slightly larger than the frequency corresponding to
the Fabry-Perot resonance, the reflection and transmission
coefficients do not have a resonant behavior. This is shown in Fig.
\ref{p01234}, where the reflection and transmission coefficients are
plotted for frequencies larger than the design frequency by $1, 2$
and 4\%. This means that the slab does not support guided modes in
this frequency range. Further increase of the frequency (by $8$ or
16\%, for example, see Fig. \ref{p01234}) leads to the increase of
the width of the narrow dip of the transmission characteristic at
$k_y=k$, and to the appearance of resonances in the vicinity of this
point. These resonances (shown with greater detail in Fig.
\ref{p0481216}) are also related with guided modes. The dispersion
branch of these modes appears near the light line in Fig.
\ref{disp}.
The resonant behavior in the present case has a very narrow band
character and does not significantly affect the imaging properties
of the device.

In Fig. \ref{p01234} one can see a strong reduction of the
reflection coefficient for $kd/\pi=1.04$ and $kd/\pi=1.08$. The
reflection coefficient becomes less than 10\% practically for the
whole spatial spectrum. This fact has been numerically confirmed,
and means that for these frequencies there is little interference
between the source and the signal reflected at the wire medium slab.

\section{Study of resolution}

In order to study the resolution of the wire medium slab, we use the
Rayleigh criterion: the resolution is taken equal to the radius of
the image spot (at the half-intensity level) for a very sharp
(nearly point) source. The resolution value obtained using this
criterion corresponds to the half of the minimum distance between
two sharp maxima that can be resolved using the imaging device.

\begin{figure}[htb]
\centering \epsfig{file=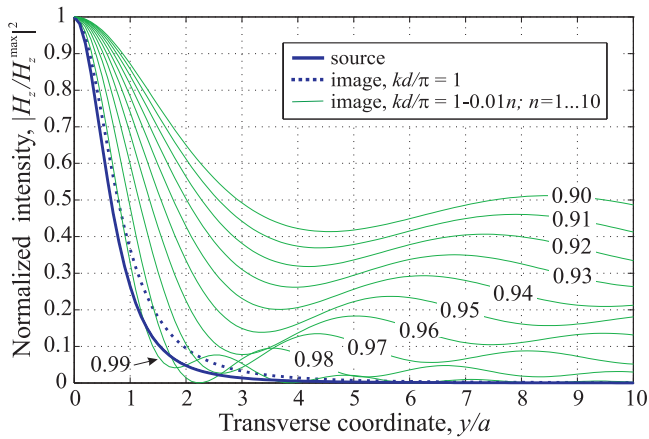, width=8.5cm} \caption{(Color
online) Image when the wire medium slab is illuminated with the
source \r{source} at frequencies slightly smaller than $kd/\pi=1$.
The numbers in the figure correspond to the values of the normalized
frequency $kd/\pi$.} \label{im1}
\end{figure}

In order that the wire medium homogenization model can be used we
are not allowed to consider any sources with radius much smaller
than the real period of the structure. Indeed the spatial spectrum
of such a source contains important spatial harmonics in the range
$k_y a\gg\pi$, whose propagation can not be described by homogenized
model. Moreover, it was shown in \cite{Subwavelength} that the
resolution of any periodic structure is limited by its period. Due
to these reasons, below we consider a magnetic field distribution
with diameter only slightly smaller than the period. The spectrum of
the source at the front plane of the wire medium slab is, \e
s(k_y)=e^{-\sqrt{k_y^2-k^2}a/2}. \label{eq:spectrum}\f Note that the
spectrum is independent of $k_z$, and consequently the field
intensity is uniform along the $z$-direction. This two-dimensional
source has the magnetic field distribution: \e
H_z(y)=A\int\limits_{-\infty}^{+\infty}s(k_y)e^{-jk_yy}dk_y
\label{eq:source}\f
$$
=Aj\pi
H_1^{(2)}\left(k\sqrt{(a/2)^2+y^2}\right)\frac{a/2}{\sqrt{(a/2)^2+y^2}},
$$
where $H_1^{(2)}(x)$ is the Hankel function of the second kind, and
$A$ is a constant that defines the amplitude of the source. The
intensity of the field normalized by its maximum value $H_z^{\rm
max}=H_z(0)$ is plotted in Fig. \ref{im1} (thick solid line).

The distribution of magnetic field under consideration appears at
the half-period distance from a plane with point source of magnetic
field which has uniform spatial spectrum. Thus, expression
\r{spectrum} also represents the transmission coefficient for a slab
of free space with the half-period thickness. This slab allows to
transform a singular distribution of the point source into less
sharp distribution which is appropriate for our studies using
effective medium theory. Note, that the point source of magnetic
field we are dealing with has nothing to do with magnetic line
source (line of magnetic current flowing along $z$-direction)! The
point source of magnetic field is a delta function of magnetic field
in the plane $x=-a/2$ [$H_z(y)|_{x=-a/2}=\delta(y)$] which means
that the field in that plane is zero everywhere except at one
singular point. We have chosen this source because of the simplicity
of its exponentially decaying spatial spectrum \r{spectrum} and
since its field decays faster than the field created by a point
magnetic source: the field of our source \r{source} decays inversely
proportional to $y$, but the field radiated by a magnetic line
source would decay as the inverse square root of $y$.

The magnetic field at the output plane when the slab is illuminated
by the source is given by: \e
H_z(y)=A\int\limits_{-\infty}^{+\infty}s(k_y)T(k_y)e^{-jk_yy}dk_y.\label{eq:image}\f

\begin{figure}[htb]
\centering \epsfig{file=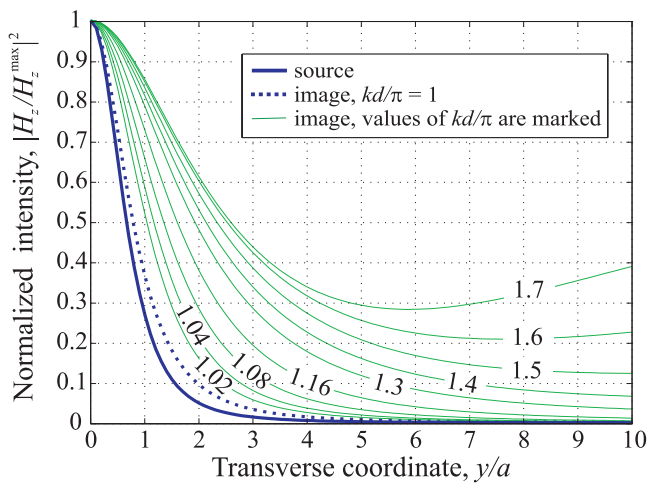, width=8.5cm} \caption{(Color
online) Image when the wire medium slab is illuminated with the
source \r{source} at frequencies slightly larger than $kd/\pi=1$.
The numbers in the figure correspond to the values of the normalized
frequency $kd/\pi$.} \label{im2}
\end{figure}

The integral in \r{image} is singular and non-integrable if the slab
supports guided modes: some poles of the transmission coefficient
\r{T} as a function of $k_y$ are located at the real line (see the
Appendix). These difficulties can be avoided if negligibly small
losses are introduced into the permittivity of the host medium. The
presence of losses shifts the poles away from the real line. This
makes the integral non-singular and it can be relatively easy
evaluated using direct numerical integration. The integral \r{image}
has been calculated for all values of $kd/\pi$ considered in the
previous section. The corresponding images at the output plane are
plotted in Fig. \ref{im1} and \ref{im2} for frequencies below and
above the first Fabry-Perot resonance, respectively. For $ka<0.93$
and $ka>1.5$ the image is distorted by the resonant guided modes of
the slab with short and long wavelengths, respectively. On the other
hand, for $0.93<ka/\pi<1.5$ the images have the same shape as the
source and their radii vary from $a$ to $2a$. This means that at
these frequencies the resolution of the imaging device $\Delta$ is
from $\lambda/30$ to $\lambda/15$. Note, that
$\Delta_l=\lambda/30=a$ is the ultimate limit of resolution imposed
by the periodicity of the structure \cite{Subwavelength}. The
resolution of $\lambda/15$ for the structure under consideration has
been observed in \cite{SWIWM} both numerically and experimentally.
This allows us to conclude that the theoretical study presented in
this paper is consistent with results reported in \cite{SWIWM}.
However, the presented analysis gives a much more comprehensive
explanation of the limitations and characteristics of our imaging
device.

\section{Accuracy of imaging}

The Rayleigh criterion used in the previous section is classical for
imaging above diffraction limit. However, near-field sub-wavelength
imaging significantly differs from conventional imaging. Indeed, the
latter requires that the source is placed very close to the
interface of the imaging device. Hence, not only the transmission
properties of the structure affect its performance, but also the
reflection properties. If the reflection coefficient has large
amplitude then the reflected field can interact with the source and
modify its near field pattern. This modified pattern will be
transmitted with sub-wavelength resolution from the input plane to
the output plane, but then it has nothing to do with the original
source. As discussed in the previous section and seen in Figs.
\ref{m002468}, \ref{m0123456} and \ref{p01234}, the strong
reflections from the slab of wire medium are mainly caused by the
resonances associated with the guided modes. That is why the imaging
with minimum reflections happens at the frequencies corresponding to
the band gaps for the guided modes and slightly higher frequencies.
Fig. \ref{p01234} clearly illustrates this statement: for
$1.01<kd/\pi<1.16$ the reflection coefficient from the slab is
smaller than 20\%, and for for $1.04<kd/\pi<1.08$ it is smaller than
10\%.

Another important point concerning accuracy of the sub-wavelength
imaging using slabs of wire media is that the transmission
coefficient is different for different transverse components of the
wave vector. As it was already mentioned, even if the thickness of
the slab is tuned to fulfil Fabry-Perot condition (as in Fig.
\ref{ka1}) the transmission coefficient is close to unity only in a
limited range of spatial harmonics. For spatial harmonics with large
$k_y$ the transmission coefficient approaches $1/2$. This fact
causes distortion in the image. It is possible to estimate the
resolution at which the imaging device works with acceptable
distortion. With this purpose we use the a half-intensity criterion:
we assume that imaging with acceptable distortion happens while the
transmission coefficient of spatial harmonics is in the range
$[1/\sqrt{2},\sqrt{2}]$. If $k_y^{\rm max}$ is the maximum
transverse wave vector component for which the half-intensity
criterion is fulfilled then the resolution is $\Delta=\pi/k_y^{\rm
max}$. This definition of resolution is stronger than the Rayleigh
criterion. It guarantees not only that maxima at distance $\Delta$
are resolved, but also that the image is restored with little
distortion.

For the case presented in Fig. \ref{ka1}, the transmission
coefficient $T$ is less then 1 and greater than $1/\sqrt{2}$ for
$k_yd/\pi<5$ except for the very narrow range of $k_yd/\pi$ close to
1. This means that the resolution of the transmission device under
study is equal to $\Delta=\pi/k^{\rm max}_y=\lambda/10$. This value
is only three times larger than the ultimate limit of resolution due
to periodicity of the structure $\Delta_l=a=\lambda/30$ formulated
in \cite{Subwavelength}, and a bit worse than the real resolution of
$2a=\lambda/15$ observed at \cite{SWIWM}.

\begin{figure}[htb]
\centering \epsfig{file=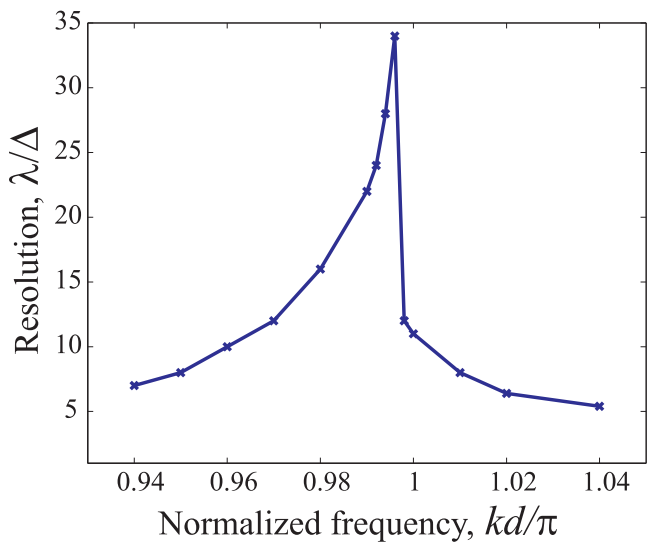, width=8.5cm} \caption{(Color
online) Normalized resolution $\lambda/\Delta$ as a function of the
normalized frequency of operation $kd/\pi$.} \label{resol}
\end{figure}

A slight change of frequency may help improving the resolution of
the device. For example, in the case of $kd/\pi=0.996$ (see Fig.
\ref{m002468}) the resolution happens to be equal to $\lambda/34$
which is even smaller than the ultimate limit dictated by
periodicity. When $kd/\pi=0.96$ the resolution is reduced down to
$\lambda/10$. One can verify that imaging with a resolution better
than $\lambda/10$ is observed for $kd/\pi\in [0.96,1.00]$. This
means that the bandwidth of operation with $\lambda/10$ or better
resolution is equal to 4\% for the transmission device under
consideration. The bandwidth of imaging with worse but still
sub-wavelength resolution can be even larger. In \cite{SWIWM} the
15\% bandwidth of imaging with sub-wavelength resolution has been
reported. We summarize our results for the dependence of the
resolution (defined using the half-intensity criterion) on the
frequency in Fig. \ref{resol}.
By tuning the frequency one can dramatically enhance the resolution.
In practice this means that it is always possible to reach an
ultimate limiting resolution $\Delta_l=a$ \cite{Subwavelength} by
appropriately choosing the frequency of operation. The value of the
resolution for the case $kd/\pi=1$ approximately describes an
average level of resolution of the system. This value can be
evaluated analytically as explained next.

From equation \r{app} it is clear that for $k_y\gg k$
(sub-wavelength spatial spectrum) then \e T\approx \mp
\frac{1}{1+\frac{k_y}{\sqrt{k_y^2+k_p^2}}}.\f Solving the equation
$T(k_y^{\rm max})=1/\sqrt{2}$ we obtain \e k_y^{\rm
max}=\sqrt{\frac{\sqrt{2}-1}{2}}k_p\approx 0.455 k_p. \f

\begin{figure}[t]
\centering \epsfig{file=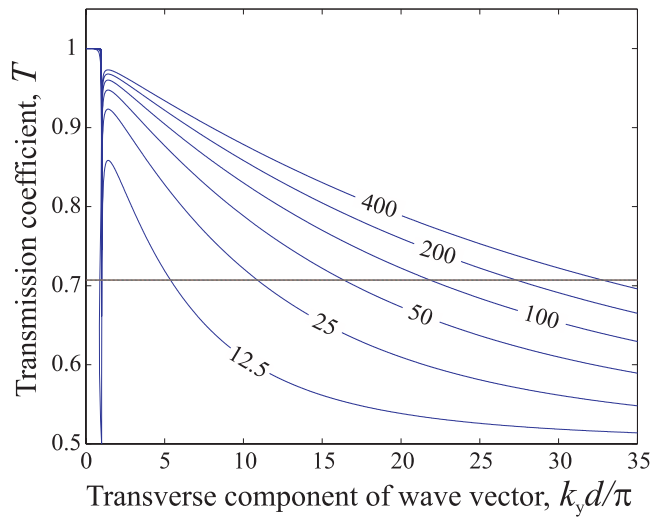, width=8.5cm} \caption{ (Color
online) Dependence of the transmission coefficient $T$ (absolute
value) on transverse component of wave vector $k_yd/\pi$ for the
case when $kd/\pi=1$ and various values of plasma frequency
$k_pd/\pi=12.5\cdot 2^n$, where n=0,1,2,3,4. The numbers at the
figure correspond to the values of $k_pd/\pi$} \label{plasma16}
\end{figure}

Thus, the resolution $\Delta=\pi/k_y^{\rm max}$ for $kd/\pi=1$ is
given by: \e \Delta= \frac{\pi}{0.455 k_p}=1.1 \lambda
\frac{k}{k_p}= 2.75 a \sqrt{\ln\frac{a}{2\pi r}+0.5275}
 \label{eq:resol}.
\f Thus, the smaller the ratio $k/k_p$ is, the better is the
resolution of the system. Formula \r{resol} shows that the
resolution of the wire medium slab is only limited by the ability to
fabricate very dense arrays of wires, i.e. the limit of resolution
only depends ultimately on the value of the lattice constant of the
crystal. By decreasing the period of wire medium one can greatly
improve resolution of the system. This fact is illustrated in Fig.
\ref{plasma16}, which shows the transmission characteristic
calculated for different values of the the plasma frequency. For
extremely thin wires and very high frequencies the effect of losses
may not be negligible. Nevertheless, at the microwave domain, this
effect is expected to be of second order.

\section{Conclusion}

In this paper, the resolution of sub-wavelength transmission devices
formed by the slabs of wire media was studied. It was shown that the
resolution does not depend on the thickness of the structure, which
can in principle be made as thick as required by an application. By
slightly tuning the frequency of operation, it is possible to
achieve the ultimate limit on resolution dictated by the periodicity
of the system (lattice constant of the crystal), even though this
regime is narrow band. The average level of resolution ultimately
depends on the period of the lattice. By reducing the lattice
constant it may be possible to realize imaging systems with
virtually no limit of resolution. This makes the slab of the wire
medium a unique imaging device capable of transmitting a
distribution of TM-polarized electric field with sub-wavelength
resolution at the microwave frequency range.

\section*{Acknowledgement}

The authors would like to thank Prof. Constantin Simovski from St.
Petersburg State University of Information Technologies, Mechanics
and Optics (Russia) for useful discussions of the results presented
in this paper.

\appendix
\section{Guided modes in the slab of wire medium}

A closely-spaced chain of half-wavelength wires \cite{Yagi}
(one-dimensional Yagi antenna array) can be regarded as a waveguide
formed by resonant scatterers. This structure is in a certain sense
the microwave analogue of optical plasmonic waveguides
\cite{Maier1,Maier2,Weber}. A slab of wire medium can be considered
as a two-dimensional analogue of the one-dimensional Yagi antenna
array. The eigenmodes of such a waveguide were studied in
\cite{Nefedovwaveguide}, but the wire medium was modeled as a local
dielectric. Below, we derive the dispersion equation for the
eigenmodes taking into account both the spatial dispersion effects
\cite{WMPRB}, and the additional boundary condition proposed in
\cite{MarioABC}.

Let us consider a slab of wire medium with thickness $d$ (see Fig.
\ref{slab}), and investigate if this structure can support guided
modes travelling along the $y$-direction with some propagation
constant $q$. A guided mode must have TM polarization with respect
to the wires, since the wire medium is transparent to the other
polarization. Then, the total magnetic field (directed along the
$z$-axis) can be written as in \r{H}, but without the term
corresponding to the incident wave since the guided mode is only
supported by its own field. The application of the boundary
conditions formulated in \cite{MarioABC} (the continuity of all
components of both magnetic and electric fields at the interface
between free space and the wire medium and, consequently, the
continuity of the magnetic field and of its both first and second
derivatives by $x$) at both interfaces $x=0$ and $x=d$ provides a
homogeneous system of equations similar to \r{syst} (the system is
homogeneous due to the absence of the incident wave). A non-trivial
solution of the system exists only if the determinant of the matrix
in \r{syst} vanishes. This yields the following dispersion equation:
$$
\left[1+\frac{\gamma_{\rm TM}k_y^2 \tanh\left(\frac{\gamma_{\rm
TM}d}{2}\right)-
kk_p^2\tan\left(\frac{kd}{2}\right)}{\gamma_x(k_y^2+k_p^2)}\right]
$$
\e \times \left[1+\frac{\gamma_{\rm TM}k_y^2{\rm
ctanh}\left(\frac{\gamma_{\rm TM}d}{2}\right)+kk_p^2{\rm
ctan}\left(\frac{kd}{2}\right)}{\gamma_x(k_y^2+k_p^2)}\right]=0.
\label{eq:disp} \f

\begin{figure}[htb]
\centering \epsfig{file=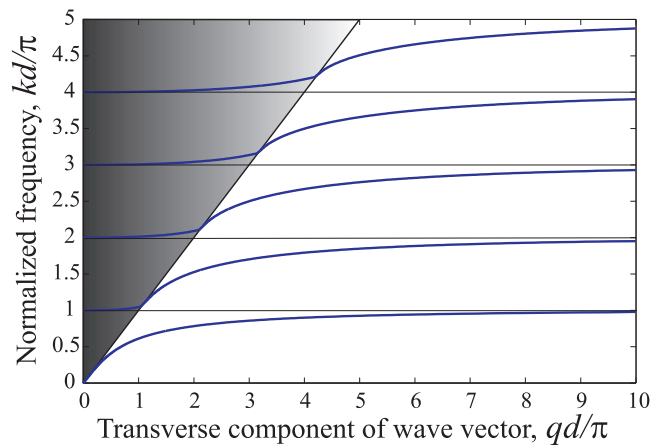, width=8.5cm} \caption{ (Color
online) Dispersion diagram for the wire medium waveguide. The region
above the light line ($q=k$) corresponding to the leaky wave region
is shadowed.} \label{disp}
\end{figure}

The dispersion equation \r{disp} is transcendental, and so cannot be
solved analytically. Its numerical solution is presented in Fig.
\ref{disp} in the form of a dispersion diagram. The parameters of
the wire medium slab are the same as in the whole paper: $a=1$~cm,
$r=1$~mm and $d=15$~cm ($k_pd/\pi=11.3$). One can see that the slab
of wire medium supports guided modes (when $q$ is a real number and
$q>k$) nearly for all frequencies, except for very narrow bands that
occur near resonances that correspond to an integer number of
half-wavelengths across the slab ($d=n\lambda/2$, where $n$ is an
integer). In this case the eigenmodes are leaky waves (${\rm
Im}(q)\ne 0$ and ${\rm Re}\{q\}<k$). The dispersion curves for the
leaky modes are presented in the shadowed region in Fig. \ref{disp}.
Only the real part of the propagation constant $q$ is shown for the
leaky modes.

The peaks of the transmission coefficient observed in Figs.
\ref{m0123456} and \ref{p0481216} are caused by the resonant
excitation of the guided modes described in this Appendix.
Mathematically, this effect can be readily explained, because when
the dispersion equation \r{disp} is fulfilled the determinant of the
matrix in \r{syst} vanishes, and consequently the reflection and
transmission coefficients \r{R}, \r{T} have poles. Thus, when an
evanescent plane wave with transverse component of the wave vector
$k_y$ equal to the propagation constant $q$ of the guided mode
illuminates the slab, the transmitted wave is very much amplified
because of the described resonant phenomenon. Note, that in Fig.
\ref{p01234} the resonances are absent for $1\le kd/\pi\le 1.06$
since the guided modes do not exist in the considered frequency
range and the leaky modes cannot be excited.

\bibliography{resolution}
\end{document}